\documentclass[twocolumn,nofootinbib,amsmath,amssymb,a4paper]{revtex4}

\usepackage{graphicx}
\usepackage{dcolumn}
\usepackage{bm}
\usepackage{epsfig}

\def\ifm#1{\relax\ifmmode#1\else$#1$\fi}

\newcommand{\ks}{\ensuremath{K_S}}
\newcommand{\kl}{\ensuremath{K_L}}
\newcommand{\DKSpippim}{\ensuremath{K_S\rightarrow\pi^+\pi^-}}
\newcommand{\DKSpienu}{\ensuremath{K_S\rightarrow\pi e\nu}}
\newcommand{\DKSpiopio}{\ensuremath{K_S\rightarrow\pi^0\pi^0}}

\def\BR{{\rm BR}}

\def\pt#1,#2,{\ifm{#1\x10^{#2}}}

\def\Journal#1#2#3#4{{\it #1} {\bf #2}, #3 (#4)}

\begin{document}

\title{KLOE results on kaon decays and summary status of $V_{us}$}

\author{T. Spadaro}
 \email{tommaso.spadaro@lnf.infn.it}
\affiliation{%
Laboratori Nazionali di Frascati dell'INFN\\
Via E. Fermi, 40 00044 Frascati (Roma) Italia
}%

\begin{abstract}
Recent KLOE measurements allowing the extraction of the $V_{us}$ element of the CKM matrix are here briefly described. 
The status of the resulting value of $V_{us}$ is summarized. 
The perspectives for the completion of ongoing analyses are discussed, with particular emphasis on the measurements of
scalar form factor slopes from study of $K_{L\mu3}$ and of $V_{us}$ from the decay width of $K^{\pm}_{l3}$.
\end{abstract}

\maketitle

\section{$V_{us}$ extraction from semileptonic kaon decays}
The most precise test of the unitarity of the CKM matrix can be performed
from its first row. Letting $\Delta=|V_{ud}|^2 + |V_{us}|^2 + |V_{ub}|^2 - 1$, an accuracy of few parts in $10^{-4}$ on $\Delta$ can be reached.
The contribution of $|V_{ub}|^2$ is negligible~\cite{PDG}; 
the determination of $V_{ud}$ from super-allowed nuclear beta decays gives an uncertainty of $5\times10^{-4}$ on $\Delta$ (see Hardy's contribution in this volume), and
a similar accuracy can be reached by extracting $|V_{us}|$ from the rates $\Gamma$ for semileptonic kaon decays:
\begin{eqnarray}
\Gamma^i(K_{e3(\gamma),\,\mu3(\gamma)})&=&|V_{us}|^2\frac{C_i^2 G^2 M^5}{128\pi^3} S_{\rm
EW}\:|f^{K^0}_+(0)|^2 \nonumber \\
 & & I^{i}_{e3,\,\mu3}\: 
  (1+\delta_{e3,\,\mu3}^{i}),\nonumber
\end{eqnarray}
where $i$ indexes $K^0\to\pi^-$ and $K^+\to\pi^0$ transitions for which $C_i^2 =1$ and 1/2, respectively, $G$ is the Fermi
constant, $M$ is the appropriate kaon mass, and $S_{\rm EW}$ is a
universal short-distance electroweak correction~\cite{as}. 
The $\delta^i$ term accounts for long-distance radiative corrections depending on the meson charges and lepton masses and, for $K^\pm$,
for isospin-breaking effects. These corrections are presently known at the few-per-mil level~\cite{aa}.
The $f^{K^0}_+(0)$ form factor parametrizes the vector-current transition $K^0\to\pi^-$ at zero momentum transfer $t$, while the dependence of
vector and scalar form factors on $t$ enter into the determination of the integrals $I_{e3,\,\mu3}$
of the Dalitz-plot density over the physical region. Since $f_+$ is dominated by the vector $K\pi$ resonances, the
closest being the $K^*(892)$, the natural form for its dependence on $t$ is:
\begin{equation}
f_+(t) \propto \frac{M_V^2}{M_V^2-t},
\label{eq:pole}
\end{equation}
but it is also customary to expand the form factor in powers of $t$ up to first or second orders, as
\begin{equation}
f_+(t)\propto 1+\lambda_+\frac{t}{m^2_{\pi^+}}\mbox{ or }
1+\lambda^\prime_+\frac{t}{m^2_{\pi^+}}+ \frac{\lambda^{\prime\prime}_+}{2}\left(\frac{t}{m^2_{\pi^+}}\right)^2.
\label{eq:linquad}
\end{equation}
For the scalar form factor a linear parametrization is typically used:
\begin{equation}
f_0(t)\propto 1+\lambda_0\frac{t}{m^2_{\pi^+}}.
\label{eq:scalarslope}
\end{equation}
The difference of $f_+(0)$ from unity reflects $SU(3)$- and $SU(2)$-breaking corrections and is evaluated from purely theoretical calculations. The
reader can refer to Sachrajda's and Portol\'es's contributions in this volume for recent updates, while we will use the old Leutwyler and Roos evaluation,
\begin{equation}
f_+(0)=0.961(8), 
\label{eq:fzerolr}
\end{equation}
in the following.
The experimental inputs in the above formulae are the semileptonic decay widths, 
evaluated from the $\gamma$-inclusive BR's and from the lifetimes, and the
parameters describing the $t$-dependence of the vector and scalar form factors.
Results from KLOE measurements of all these inputs are reported in the following.

\section{Experimental setup}
DA$\Phi$NE, the Frascati $\phi$ factory, is an $e^{+}e^{-}$ collider
working at $\sqrt{s}\sim m_{\phi} \sim 1.02$~GeV. $\phi$ mesons are produced,
essentially at rest, with a visible cross section of $\sim$~3.1~$\mu$b
and decay into $\ks\kl$ ($K^+K^-$) pairs with a BR of $\sim 34$\% ($\sim 49$\%).

Kaons get a momentum of $\sim$~100~MeV/$c$ which translates into a low speed, $\beta_{K} \sim$ 0.2.
$\ks$ and $\kl$ can therefore be distinguished by their mean decay lengths:
$\lambda_{S} \sim $~0.6~cm and $\lambda_{L} \sim $~340~cm.
$K^+$ and $K^-$ decay with a mean length of $\lambda_\pm\sim $~90~cm and can be distinguished from their decays in flight to one of the two-body final states 
$\mu\nu$ or $\pi\pi^0$.

The kaon pairs from $\phi$ decay are produced in a pure $J^{PC}=1^{--}$ quantum state, so that 
observation of a $\ks$ ($K^+$) in an event signals, or tags, the presence of a $\kl$ ($K^-$)
and vice versa; highly pure and nearly monochromatic $\ks$, $\kl$, and $K^\pm$
beams can thus be obtained and exploited to achieve high precision in the measurement of absolute BR's.

The analysis of kaon decays is performed with the KLOE detector, consisting essentially of a drift chamber, DCH, surrounded by an
electromagnetic calorimeter, EMC. A superconducting coil provides a 0.52~T magnetic field.
The DCH~\cite{nimdch} is a cylinder of 4~m in diameter
and 3.3~m in length, which constitutes a fiducial volume 
for $\kl$ and $K^\pm$ decays extending for $\sim0.4\lambda_{L}$ and $\sim1\lambda_\pm$, respectively.
The momentum resolution for tracks 
at large polar angle is $\sigma_{p}/p \leq 0.4$\%. 
The invariant mass reconstructed from the momenta of the two pion tracks of a \DKSpippim\ decay peaks
around $m_K$ with a resolution of $\sim$800~keV, thus allowing clean \kl\ tagging. 
The c.m.\ momenta reconstructed from identification of 1-prong $K^\pm\to\mu\nu,\pi\pi^0$ decay vertices in the DC 
peak around the expected values with a resolution of 1--1.5~MeV, thus allowing clean and efficient $K^\mp$ tagging. 

The EMC is a lead/scintillating-fiber sampling calorimeter~\cite{nimcalo}
consisting of a barrel and two endcaps, with good
energy resolution, $\sigma_{E}/E \sim 5.7\%/\sqrt{\rm{E(GeV)}}$, and excellent 
time resolution, $\sigma_{T} =$~54~ps$/\sqrt{\rm{E(GeV)}} \oplus 50$ ps. About 50\% of the \kl's produced reach the EMC, where most interact.
A signature of these interactions is the presence of  
an high-energy cluster not connected to any charged track, with a time corresponding to a low velocity: 
the resolution on $\beta_K$ corresponds to a resolution of $\sim1$~MeV
on the \kl\ momentum. This allows clean \ks\ tagging.
The timing capabilities of the EMC are also exploited to precisely reconstruct the position of decay vertices of $K_L$ and $K^\pm$ to $\pi^0$'s from the
cluster times of the emitted photons, thus allowing precise measurements of the $K_L$ and $K^\pm$ lifetimes.

In early 2006, the KLOE experiment completed data taking, having collected
$\sim2.5$~fb$^{-1}$ of integrated luminosity at the $\phi$ peak,
corresponding to $\sim$2.5~(3.6) billion $K_S K_L$ ($K^+ K^-$) pairs. The results 
presented here are based on the first 400~pb$^{-1}$ collected and are based
on analyses published in 2006.
\section{$V_{us}$ from semileptonic \kl\ decays}
\subsection{Measurements of $K_{Le3}$ and $K_{L\mu3}$ BR's}
The analysis of \kl\ decays starts with the identification of \DKSpippim\ decays, which gives a pure \kl\ ``beam'' of known momentum and direction. 
In a fiducial volume extending
for $\sim0.4\lambda_L$, two-track decay vertices are selected around the \kl\ line of flight and the number of events for
each of the decay modes $K_L\to \pi e\nu,$ $\pi\mu\nu,$ and $\pi^+\pi^-\pi^0$ are obtained from the distribution of the difference $E_{\rm miss}-P_{\rm miss}$ of
missing momentum and missing energy in the hypotheses of pion and muon daughter particles.
Photon vertices from $\kl\to3\pi^{0}$ decays are reconstructed on the \kl\ line of flight from the times of at least 3 photon clusters. Since the geometrical
acceptance of these selections depends on the value of the \kl\ lifetime, the output
values of the BR's are expressed as 
a function of $\tau_L$~\cite{KLOE+06:KLBR}:
\begin{eqnarray}
\BR(K_L\to\pi e\nu)       &=&\frac{40.49(10)_{\rm st}(18)_{\rm sy}\%}{1+k\Delta\tau}\\
\BR(K_L\to\pi\mu\nu)      &=&\frac{27.26(9)_{\rm st}(14)_{\rm sy}\%}{1+k\Delta\tau}\\
\BR(K_L\to\pi^0\pi^0\pi^0)&=&\frac{20.18(5)_{\rm st}(23)_{\rm sy}\%}{1+k\Delta\tau}\\
\BR(K_L\to\pi^+\pi^-\pi^0)&=&\frac{12.76(6)_{\rm st}(14)_{\rm sy}\%}{1+k\Delta\tau},
\end{eqnarray}
where $\Delta\tau=51.7\mbox{~ns}-\tau_{L}$, $k=0.0128$~ns$^{-1}$, and the uncertainties on the above results are correlated with the following coefficients:
\begin{equation}
\left(
\begin{array}{cccc}
1& 0.09& 0.07 & 0.49 \\
 & 1   &-0.03 & 0.27 \\
 &     & 1    & 0.07 \\
 &     &      & 1
\end{array}\right)
\end{equation}
The above inputs are to be used for the evaluation of world-average \kl\ BR's.
Imposing the unitarity of the above ratios, 
$\sum_{i}BR_{i} = 1-\BR(K_L\to\pi\pi)-\BR(K_L\to\gamma\gamma)= 1 - 0.36\%$, we can extract both the four main BR's and $\tau_L$. This calculation is handled by performing a fit
to the above measurements, together with the direct KLOE measurement of $\tau_L$ from 
the $\kl\to\pi^0\pi^0\pi^0$ decay distribution, 
$\tau_L=50.92(17)_{\rm st}(25)_{\rm sy}$~ns~\cite{KLOE+05:KLlife}.
The results are
\begin{eqnarray}
\BR(K_L\to\pi e\nu)       &=&40.08(6)_{\rm st}(14)_{\rm sy}\%\\
\BR(K_L\to\pi\mu\nu)      &=&26.99(6)_{\rm st}(13)_{\rm sy}\%\\
\BR(K_L\to\pi^0\pi^0\pi^0)&=&19.96(5)_{\rm st}(19)_{\rm sy}\%\\
\BR(K_L\to\pi^+\pi^-\pi^0)&=&12.61(5)_{\rm st}(10)_{\rm sy}\%\\
\tau_L            &=&50.84(14)_{\rm st}(18)_{\rm sy}\mbox{ ns,}
\end{eqnarray}
with correlation matrix
\begin{equation}
\left(
\begin{array}{ccccc}
1& -0.31 & -0.55 & -0.01 &  0.16 \\
 &  1    & -0.41 & -0.14 &  0.22 \\
 &       &  1    & -0.47 & -0.14 \\
 &       &       &  1    & -0.26 \\
 &       &       &       &  1    
\end{array}\right).
\end{equation}
\subsection{\kl\ form factor slopes}
From the sample of charged \kl\ decays, additional loose cuts on kinematics and improved particle identification 
from the time of flight (TOF) of daughter particles 
(evaluated from connected EMC clusters) 
allow the selection of a high-purity $2\times10^6$-event sample of $K_L\to\pi^{\mp}e^{\pm}\nu(\overline{\nu})$ decays. Within this sample,
the probability of misidentifying an electron as a pion is negligible, so that the momentum tranfer $t$ can be safely evaluated from the \kl\ momentum and from the
momenta of the daughter tracks. The vector form factor slopes are extracted through binned log-likelihood fits of $t$ distributions to the parametrizations
of Eqs.~\ref{eq:pole} and~\ref{eq:linquad}. The results are~\cite{KLOE+06:KLe3FF}:
\begin{eqnarray}
                  M_V & = & (870\pm6_{\rm st}\pm7_{\rm sy})\mbox{ MeV,} \\
\lambda_+             & = & (28.6\pm0.5_{\rm st}\pm0.4_{\rm sy})\times10^{-3},
\end{eqnarray}
and
\begin{equation}
\begin{array}{ccc}
\lambda^\prime_+       & = & (25.5\pm1.5_{\rm st}\pm1.0_{\rm sy})\times10^{-3}\\
\lambda^{\prime\prime}_+ & = & ( 1.4\pm0.7_{\rm st}\pm0.4_{\rm sy})\times10^{-3}\\
\rho(\lambda^\prime_+,\lambda^{\prime\prime}_+) & = & -0.95
\end{array}
\label{eq:ke3ff}
\end{equation}
The KLOE inputs from $K_{Le3}$ decays allow the evaluation of $f_+(0)\times|V_{us}|=0.21561(69)$, i.e., with an accuracy of 0.32\%.

The analysis of a $K_{L\mu3}$ sample for the measurement of the scalar form factor slope, Eq.~\ref{eq:scalarslope},
 is more complicated than for $K_{Le3}$ because pure and
efficient $\pi$-$\mu$ separation is much more difficult to achieve. In order to overcome this problem, an analysis presently in progress aims at measuring
$\lambda_0$ through a fit to the distribution of the neutrino energy $E_\nu$, which can be evaluated simply through a Lorentz transform of $P_{\rm miss}$ 
in the \kl\ frame. As a side effect, the sensitivity on $\lambda_{+}$ ($\lambda_0$) from a $E_\nu$ fit is reduced by a 
factor of $\sim$2 (1.25) with respect to that achieved from a $t$ fit. The relative statistical accuracy on $\lambda_0$ will be in the range of 5--10\% after
analysis of the entire data set.

\begin{figure}[ht]
\centerline{\psfig{file=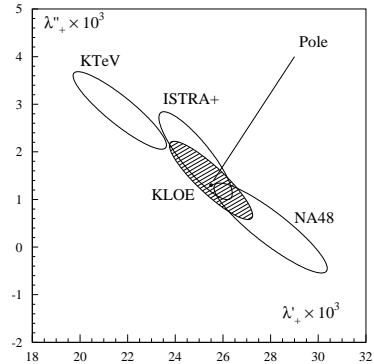,width=2.in}}
\caption{Measurements of $\lambda_+'$ and $\lambda_+''$: 1-$\sigma$ contours are shown for each experiment. The 
dot shows ($\lambda_+'$, $\lambda_+''$) values corresponding to the pole-fit result of KLOE data.}
\label{fig:ff}
\end{figure}

The results from quadratic fits of $K_{Le3}$ can be compared to and combined with the measurements from $K_{Le3}$ decays from NA48~\cite{NA48+04:Ke3FF}, 
and from $K_{Le3}$ and $K_{L\mu3}$ decays from
KTeV~\cite{KTeV+04:FF}, and ISTRA+\cite{ISTRA+04:Kl3}. 
This comparison has to be performed with correlations taken into account.
Using the method of~\cite{Moulson+06:CKM06}, the values obtained are (see Fig.~\ref{fig:ff}):
\begin{equation}
\begin{array}{ccc}
\lambda^\prime_+       & = & (24.92\pm0.83)\times10^{-3}  \\
\lambda^{\prime\prime}_+ & = & ( 1.59\pm0.36)\times10^{-3}\\
\lambda_0             & = & (16.07\pm0.82)\times10^{-3},
\end{array}
\label{eq:kl3ff}
\end{equation}
with $\chi^2/{\rm dof} = 11.9/9$ (21.7\%) and correlation matrix
\begin{equation}
\begin{array}{ccc}
\rho(\lambda^\prime_+,\lambda^{\prime\prime}_+) & = & -0.94 \\
\rho(\lambda^\prime_+,\lambda_0)             & = &  0.24    \\
\rho(\lambda^{\prime\prime}_+,\lambda_0)       & = & -0.34.
\label{eq:combinedslopes}
\end{array}
\end{equation}
The above result, together with the KLOE measurements of $\BR(K_L\to\pi\mu\nu)$ and $\tau_L$, gives $f_+(0)\times|V_{us}|=0.21633(78)$, i.e., with an accuracy of 0.36\%.
\section{$V_{us}$ from semileptonic \ks\ decays}
The analysis of \ks\ decays starts with the identification of \kl\ interactions in the EMC, which gives a \ks\ ``beam'' of known momentum and direction. 
Two-track decay vertices close to the interaction point are selected; in order to reject the background from \DKSpippim\ decays, we apply a PID technique
exploiting the time of the clusters connected to the \ks\ daughter tracks. After this selection, pion and electron tracks are precisely identified. 
About 10\,000 events for each of the $K_S\to \pi^{+}e\overline{\nu}$ and $\pi^-e^+\nu$ decay modes are obtained from the distribution of the difference 
$E_{\rm miss}-P_{\rm miss}$ of missing momentum and missing energy. By normalizing the number of $K_{e3}$ counts to the number of \DKSpippim\ counts in the same date set
and correcting for the efficiency ratio for each charge state, we obtain~\cite{KLOE+06:KSe3}:
\begin{equation}
\begin{array}{ccccc}
R_{+} & = & \frac{\Gamma(\ks\to\pi^-e^+\nu)}{\Gamma(\DKSpippim)} &=& 5.099(82)_{\rm st}(39)_{\rm sy}\times10^{-3}\\
R_{-} & = & \frac{\Gamma(\ks\to\pi^+e^-\overline{\nu})}{\Gamma(\DKSpippim)} &=& 5.083(73)_{\rm st}(42)_{\rm sy}\times10^{-3}\\
\end{array}
\end{equation}
The total errors are $\sim1\%$ and are dominated by the statistics for signal events and background subtraction. 
Combining $R_{\pm}$ with the precise KLOE measurement of the ratio~\cite{KLOE+06:KSratio},
\begin{equation}
R   = \frac{\Gamma(\DKSpippim)}{\Gamma(\DKSpiopio)} = 2.2549\pm0.0054,
\end{equation}
and imposing $e$-$\mu$ universality, we obtain the main $K_S$ BR's:
\begin{eqnarray}
\BR(\DKSpippim)                   &=&(69.196\pm0.051)\times 10^{-2}\\
\BR(\DKSpiopio)                   &=&(30.687\pm0.051)\times 10^{-2}\\
\BR(\ks\to\pi^-e^+\nu)            &=&(3.528\pm0.062) \times 10^{-4}\\
\BR(\ks\to\pi^+e^-\overline{\nu}) &=&(3.517\pm0.058) \times 10^{-4},
\end{eqnarray}
from which we obtain $\BR(\DKSpienu)=(7.046\pm0.091)\times10^{-4}$.
These are by far the most precise measurements of these BR's and allow a substantial improvement in the accuracy of the determination of $CP$ and $CPT$ parameters of the $K^0$ system using the Bell-Steinberger relation~\cite{KLOE+06:BSR}.
An enriched sample of $\sim$10\,000 $K_{Se3}$ events with a 0.7\% overall
contamination can be obtained by tightening the kinematic cuts. From this, a linear fit to the $t$-distribution yields the first measurement ever made
of the vector form factor slope $\lambda_+$ for $K_{Se3}$:
\begin{equation}
\lambda_+              =  (33.9\pm4.1)\times10^{-3}.
\end{equation}
Using the much more precise determination of the slopes from $K_{Le3}$ decays and the precise average value of 
$\tau_{S}$ from the PDG~\cite{PDG}, our value for $\BR(\DKSpienu)$ gives
$f_+(0)\times|V_{us}|=0.2154(14)$, i.e., with an accuracy of 0.67\%. This determination is unique to KLOE.

The event sensitivity at KLOE to $K_{S\mu3}$ decays is lower than that for $K_{Se3}$. The reasons are similar to those presented for the analysis of the
$K_{L\mu3}$ form factor slopes. In addition, in this case the background is dominated by events having the same particles as the signal:
in \DKSpippim\ decays, a $\pi$ may decay to $\mu\nu$ before entering the DCH. The BR for $K_{S\mu3}$ decay has never been
measured before. The analysis is in progress and a total accuracy of 3\% on the BR should be obtained after the entire data set is analyzed.

\section{$V_{us}$ from semileptonic $K^+$ decays}
\subsection{Measurements of $K^\pm_{e3,\mu3}$ BR's} 
The analysis of $K^\pm$ decays starts with the identification of $K^\mp$ decays to $\mu^\mp\overline{\nu}(\nu)$ or $\pi^\mp\pi^0$ final states, 
which gives a pure sample of $K^\pm$-tagged events.
One-prong decay vertices of $K^\pm$ are then selected in the DCH and the photons coming from $\pi^0$ decay are identified from TOF using the associated clusters in the EMC.
In order to reject the background for $K^\pm_{e3,\mu3}$ identification, which is dominated by $\pi^+\pi^0$ and $\tau^\prime$ decays, the lepton TOF is evaluated
exploiting the time of the cluster connected to the $K^+$ daughter track. 
The number of events for each of the $K^+\to \pi e\nu$ and $\pi\mu\nu$ decay modes are obtained from the distribution of the squared lepton mass evaluated from TOF.
By normalizing to the number of tagged events and correcting for the selection efficiency, 
we obtain the following preliminary results~\cite{KLOE+05:lisbonproc}:
\begin{eqnarray}
\BR(K^\pm\to e^\pm\pi^0\nu(\overline{\nu})) &=& 5.047(39)_{\rm st}(81)_{\rm sy}\%\\
\BR(K^\pm\to\mu^\pm\pi^0\nu(\overline{\nu}))&=& 3.310(45)_{\rm st}(65)_{\rm sy}\%\\
\rho(\BR(K^\pm_{e3}),\BR(K^\pm_{\mu3}))     &=& 0.42.
\end{eqnarray}
\subsection{Measurements of $K^\pm$ lifetime} 
The experimental status of $\tau_+$ is unclear: the PDG quotes an average of $\tau_+=12.385(25)$~ns~\cite{PDG} 
with a relative accuracy of 0.2\% and a confidence level of 0.2\%. At KLOE there are two methods to perform a direct measurement of $\tau_+$ from
the distribution of the proper decay times $t^\ast$. One can obtain $t^\ast$ from the $K^\pm$ track length in 1-prong kaon decays, properly accounting for
kaon energy loss in each track segment $L_{i}$: $t^\ast=\sum_{i}L_{i}/(\beta_{i}\gamma_{i}c)$. An independent determination of $t^\ast$ can be obtained
from $K^\pm$ decays to final states containing $\pi^0$'s, by using the photon TOF's. While this second method is still under development, a preliminary result from the first approach
has been obtained using a sample of $\sim175$~pb$^{-1}$:
\begin{eqnarray}
\tau_+            &=&12.367\pm0.044_{\rm st}\pm0.065_{\rm sy}\mbox{ ns,}
\end{eqnarray}
where the systematic uncertainty has been conservatively evaluated. After the analysis of the entire data set, KLOE results are expected to clarify the experimental situation concerning $\tau_{+}$. In the following, we will use the PDG value of $\tau_+$.
\section{KLOE summary of $f_0\times|V_{us}|$}
Using the form factor slopes (FF) from Eq.~\ref{eq:ke3ff} for $K_{e3}$ and the averages of Eq.~\ref{eq:kl3ff} for $K_{\mu3}$ modes, the values 
of $f_+\times|V_{us}|$ from KLOE measurements are:
\begin{equation}
\begin{tabular}{|l|l|l|l|l|}\hline
\multicolumn{3}{|c|}{} &  \multicolumn{2}{c|}{Input} \\ 
 Mode         & $f_+\times|V_{us}|$ & Error,\%     &  \multicolumn{1}{c}{KLOE}  & \multicolumn{1}{c|}{External} \\ \hline 
$K_{Le3}$     & 0.21561(69)       &  0.32          & FF, BR, $\tau_L$ &              \\
$K_{L\mu3}$   & 0.21633(78)       &  0.36          & FF, BR, $\tau_L$ & FF           \\
$K_{Se3}$     & 0.2154(14)        &  0.67          & FF, BR           & $\tau_S$     \\
$K^\pm_{e3}$  & 0.2170(21)        &  0.96          & FF, BR           & $\tau_+$     \\
$K^\pm_{\mu3}$& 0.2150(28)        &  1.3           & FF, BR, $\tau_L$ & FF, $\tau_+$   \\ \hline
\end{tabular}
\end{equation}
The best accuracy is obtained from $K_L$ modes, with errors dominated by $\tau_L$; 
intermediate accuracy is obtained from $K_{Se3}$, with error dominated
by the BR measurement. If the average FF's are used for each mode, the following results are obtained:
\begin{equation}
\begin{tabular}{|l|l|l|l|l|}\hline
\multicolumn{3}{|c|}{} &  \multicolumn{2}{c|}{Input} \\ 
 Mode         & $f_+\times|V_{us}|$ & Error,\%     &  \multicolumn{1}{c}{KLOE}  & \multicolumn{1}{c|}{External} \\ \hline 
$K_{Le3}$       & 0.21572(64)       &  0.30          & FF, BR, $\tau_L$ & FF            \\
$K_{L\mu3}$     & 0.21633(78)       &  0.36          & FF, BR, $\tau_L$ & FF            \\
$K_{Se3}$       & 0.2155(14)        &  0.66          & FF, BR           & FF, $\tau_S$  \\
$K^\pm_{e3}$    & 0.2171(21)        &  0.96          & FF, BR           & FF, $\tau_+$  \\
$K^\pm_{\mu3}$  & 0.2150(28)        &  1.3           & FF, BR, $\tau_L$ & FF, $\tau_+$  \\ \hline
Average         & 0.21595(50)       &  0.23          & \multicolumn{2}{c}{}             \\ \cline{1-3}
\end{tabular}
\end{equation}
The average in the last line is evaluated taking correlations into account and has a 86\% $\chi^2$ probability.
Using the Leutwyler-Roos value for $f_+(0)$ gives 
\begin{equation}
|V_{us}|=0.2247(19).
\label{KLOEVus}
\end{equation}
Using the world-average value of $V_{ud}=0.97377(27)$ as obtained from $0^+\to0^+$ nuclear beta decays~\cite{Marciano+06:vud},
we get $\Delta = (-13\pm10)\times10^{-4}$. At the time of the 2004 PDG compilation~\cite{PDG04}, the world-average value was 
 $\Delta=(-35\pm15)\times10^{-4}$, i.e., $\sim2.3\sigma$ away from zero.

The universality of $e$ and $\mu$ couplings to the $W$ demands that the values of $V_{us}$ obtained from $K_{e3}$ and $K_{\mu3}$ decays be the same. The KLOE data satisfy the $e/\mu$ universality test: unlike at the time of
2004 PDG~\cite{PDG04}, the ratios of effective Fermi constants are now compatible with unity:
\begin{equation}
\begin{tabular}{|c|c|c|}\hline
 Source & \multicolumn{2}{c|}{$G^2(\mu3)/G^2(e3)$} \\ \hline
$K_{L}$ & $1.0059(83)$                       & $1.047(14)$                         \\
$K^\pm$ & $0.981(25)$                        & $1.004(16)$                         \\ \hline
\multicolumn{1}{c|}{}        & KLOE 06 & PDG 04 \\ \cline{2-3}
\end{tabular}
\label{tab:gfmgfe}
\end{equation}
\section{KLOE contribution to $|V_{us}/V_{ud}|$}
By comparing radiation-inclusive kaon and pion widths for $\mu\nu$ decays, one can extract the ratio $|V_{us}/V_{ud}|$ from the following
relation~\cite{Marciano+04:VusVud}:
\begin{equation}
\frac{\Gamma(K\to\mu\nu)}{\Gamma(\pi\to\mu\nu)} = 
\frac{m_K\left(1-m_\mu^2/m_K^2\right)^2}{m_\pi\left(1-m_\mu^2/m_\pi^2\right)^2}
\left|\frac{V_{us}}{V_{ud}}\right|^2
\frac{f_K^2}{f_\pi^2}
C
\end{equation}
The theoretical inputs are the form-factor ratio $f_K/f_\pi$ and the radiative corrections are described by the factor $C$. We
use $f_K/f_\pi=1.208(2)(^{+7}_{-14})$ 
from lattice calculations by the MILC collaboration~\cite{MILC+06:fkfp}, and
$C=0.9930(35)$ from~\cite{Marciano+04:VusVud}.

\begin{figure}[ht]
\centerline{\psfig{file=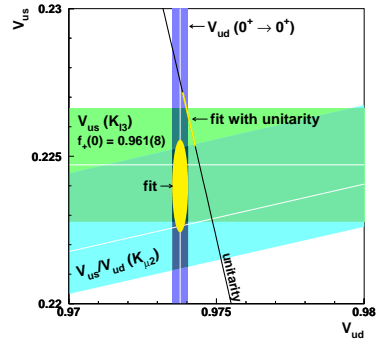,width=2.in}}
\caption{The result of a fit combining the world-average value of 
$V_{ud}$ with the KLOE measurements of $V_{us}$ and $V_{us}/V_{ud}$ 
is shown in the $V_{us}$-$V_{ud}$ plane by the solid ellipse, 
which corresponds to a 1-$\sigma$ contour. 
The unitarity constraint is also shown by the solid line. 
The segment highlighted on it represents the result of a fit 
assuming unitarity as a constraint.}
\label{fig:fits}
\end{figure}

From the precise KLOE measurement~\cite{KLOE+06:Kpmn} $\BR(K^+\to\pi^+\nu)=63.66(9)_{\rm st}(15)_{\rm sy}\%$,
and using the PDG values~\cite{PDG} for the other experimental inputs, we get $|V_{us}/V_{ud}|=0.2286(^{+27}_{-15})$.
This result can be fit together with the world-average value $V_{ud}=0.97377(27)$ and 
the $|V_{us}|$ evaluation from KLOE results, Eq.~\ref{KLOEVus}.
The fit is shown in fig.~\ref{fig:fits}. 
It yields $V_{us}=0.2240(16)$ and $\Delta=1.60(89)\times10^{-3}$ with a $\chi^{2}$ probability of 53\%,
demonstrating the consistency of the KLOE measurements, and giving no 
indication of any violation of CKM unitarity. 

\end{document}